# Information technology curriculum: General or specialized? An Australia’s census study

**Pak-Lok Poon (Corresponding Author)**
School of Engineering and Technology
Central Queensland University
Melbourne 3000, VIC, Australia
Email: p.poon@cqu.edu.au

**Sau-Fun Tang**
School of Engineering and Technology
Central Queensland University
Melbourne 3000, VIC, Australia

**Santoso Wibowo**
School of Engineering and Technology
Central Queensland University
Melbourne 3000, VIC, Australia

**Srimannarayana Grandhi**
School of Engineering and Technology
Central Queensland University
Melbourne 3000, VIC, Australia

## Abstract

Despite the strong employment prospect for information technology (IT) graduates, a comprehensive study investigating the status quo of offering different types of IT degree by Australian universities does not exist. To address this issue, this paper investigates how Australian universities offer three different types of IT degree: general, specialized, and those with majors. Using effect size analysis, we have observed some interesting phenomena about the correlation between how Australian universities offer their IT degrees and different factors, including, for example, type and reputation of universities, degree level, research component, supporting infrastructure, and industry engagement. Our census study painted the status quo of offering different types of IT degree by Australian universities, and provided insights into whether these IT degrees properly address the IT industry’s needs. Based on the findings, we have also highlighted some insights and made recommendations on how to improve IT students’ learning outcomes and graduates’ employability.

## Introduction

In Australia, the employment growth for digital workers is projected to increase up to 13.9% by 2030 (Commonwealth of Australia, 2025). Naturally, this prosperous employment outlook will attract high school leavers to apply for a Computer Science (CS) or Information Technology (IT) bachelor's study (in the rest of this paper, CS and IT will be collectively referred to as "IT"), and bachelor's degree holders to pursue a postgraduate IT study.[1] This phenomenon is supported by the 2021 Census data, showing that the number of people in Australia with an IT qualification has increased by 36% since 2016 (Australian Bureau of Statistics, 2022).

Despite a large population of IT students in universities, we observed that there does not exist a comprehensive study of IT curricula offered by Australian universities. Consequently, we do not know the status quo of IT tertiary education in Australia, and whether such education properly addresses the IT industry's needs. Thus, we conducted a census study to investigate the following three different types of Australian IT degree:

- General IT degree (GL-degree; "GL" denotes "General"): It provides broad knowledge across various IT areas such as information processing, systems analysis and design, and introductory computer programming. Examples of a GL-degree are Bachelor of Computing and Master of IT. Graduates of a GL-degree ("generalists") will be exposed to more versatile career options and adaptability. Generalists can decide on a specific IT career later, because their broad IT knowledge provides a strong foundation for specialization in various IT areas such as software development and cybersecurity.
- Specialized IT degree (SP-degree; "SP" denotes "Specialized"): It offers in-depth, focused training in specific, high-demand areas such as software engineering and artificial intelligence (AI). Examples of a SP-degree are Bachelor of Software Engineering and Master of AI. Essentially, a SP-degree equips graduates ("specialists") with deep expertise for complex, specific technical challenges. Potentially, specialists have higher "initial" career prospects in a specific IT niche due to targeted expertise.
- IT degree with majors (MA-degree; "MA" denotes "Major"): A major is a specialized IT curriculum within a broader IT degree, offering focused coursework and exposure to a specific IT field. Similar to a SP-degree, an MA-degree prepares graduates for specialized IT roles. An example of this degree is Bachelor of Computer Science (Software Engineering).[2] A major is not an academic degree by itself. This sets an MA-degree apart from a SP-degree (e.g., Bachelor of Software Engineering). In terms of curriculum outcomes, an MA-degree falls somewhere along a spectrum between a GL-degree and a SP-degree, and an MA-degree resembles more with a SP-degree than with a GL-degree.

The type of IT degree to study depends on the individual career goals, preferences, and market demands. On the one hand, students may choose a GL-degree if they prefer to acquire a versatile and adaptable skill set so that they can be exposed to diverse IT career opportunities immediately after graduation. On the other hand, students may enroll for an MA-degree or a SP-degree if they want to be job-ready and can directly enter roles that match their specialized training upon graduation.

## Underpinning theory

Curriculum theories provide the foundation for connecting educational philosophies to teaching practices. These theories help educational institutes decide what and how to teach, as well as how learning is measured (Ellingson & Roehrig, 2025). Among the various curriculum theories, one of them is *subject-centered theory* (*SCT*) or *subject-centered curriculum*, which is considered a conventional curriculum model (Yates & Collins, 2010). SCT focuses on the mastery of specific disciplines (e.g., computer science, engineering, and mathematics) (Kridel, 2010). According to SCT, within a curriculum, knowledge is categorized into distinct subjects rather than blended across disciplines (Button, 2021). SCT emphasizes transferring established knowledge from experts to learners and, hence, is primarily "instructor-directed".

Many modern IT curricula are designed based on SCT, where such curricula comprise a set of specialized IT units[3] such as computer hardware architecture, database management, data structures, systems analysis and design, and computer networking. We witnessed that different universities offer their IT curricula with different combinations of units. This difference is further complicated by offering different types of IT degrees (i.e., GL-degree, SP-degree, and MA-degree) as mentioned in the Introduction. To the best of our knowledge, we are not aware of any comprehensive study on IT curricula under the lens of SCT, particularly in an Australian context.

[1] In this paper, a postgraduate program only refers to Coursework Master's. To avoid verbosity, a Coursework Master's will simply be referred to as a Master's in this paper.

[2] When there is no ambiguity, the major of an MA-degree is enclosed within round parentheses.

[3] In this paper, a *unit* is a syllabus item offered by a university, which is similar to a subject (e.g., Mathematics and Physics) that students study at high school.

## Study settings

### *Research objectives and questions*

One research objective of this census study is to investigate the status quo of offering the three types of IT degrees by all relevant universities in Australia. To address this objective, we formulated the following research questions:

*RQ1: How many Australian universities are offering GL-degrees, MA-degrees, and SP-degrees?*

Australia has a few universities of technology (UoTs), which are more STEM-centric. All other universities are considered "comprehensive" universities (CUs), which offer bachelor's and master's programs across diverse fields, rather than solely focusing on some selective disciplines such as IT and engineering.

*RQ2: Do UoTs offer more MA-degrees and SP-degrees than CUs?*

In Australia, the Group of Eight (Go8) comprises eight research-intensive universities known for their high academic standards, significant research output, and global reputation. The Go8 universities are the Australian National University, the University of Melbourne, the University of Sydney, the University of New South Wales, the University of Queensland, the University of Western Australia, Monash University, and Adelaide University. The academic status of G08 universities in Australia is similar to the Ivy League universities in the USA and the Russell Group universities in the UK.

*RQ3: Do the Go8 universities offer more or less MA-degrees and SP-degrees than other universities?*

In Australia, regional universities are those whose are regionally headquartered, with a primary mission to serve rural/remote communities and provide equitable access to education. On the other hand, metropolitan universities are those primarily located in capital/major cities (see Table 1), offering a wide range of courses and research opportunities.

*RQ4: Do regional universities offer more or less MA-degrees and SP-degrees than metropolitan universities?*

*RQ5: Which are the most popular SP-degrees offered by Australian universities?*

*RQ6: Does the master's level offer more or less SP-degrees than the bachelor's level?*

*RQ7: Which are the most popular majors in MA-degrees?*

*RQ8: Which type(s) of IT degrees is/are more likely to incorporate a research component and/or an internship?*

*RQ9: How many Australian universities have established an associated infrastructure (in terms of professorship recruitment and research institute/center) to support teaching and research of specialized IT units?*

*RQ10: How many universities offering MA-degrees and/or SP-degrees have industrial involvement (in terms of industry advisory board, curriculum co-design, co-teaching, and internship opportunities)?*

Later in the paper, when we discuss the findings of the above research questions, we will also discuss whether the overall university IT education properly addresses the IT industry's needs.

## Participant universities and data collection

There are 42 universities in Australia, among which 37 are public universities and 5 are private ones. Our census study covered all the universities in Australia, except Australian University of Theology, University of Divinity, and Avondale University. This is because these three universities do not offer any IT study. After filtering the above three universities, 39 remained in our study. These 39 universities are called *IT-degree offering universities* or simply *offering universities*. Their geographical distributions across different regions of Australia are shown in

Table 1.[4] Information about IT studies was collected online from each offering university's website (between October–November 2025) for detailed analysis.

## Statistical analysis

### *RQ1: Number of universities offering each type of IT degrees*

Table 2 shows the numbers and percentages of universities offering each type of IT degrees. We also counted the number of universities as follows: (Group 1) those offering only GL-degrees; and (Group 2) those offering MA-degrees and/or SP-degrees (possibly with GL-degrees).

Table 3 shows the statistics. In the rightmost column of Table 3, the two totals are both 38 (not 39). This is because there is one university which does not offer any bachelor's IT degree, and there is another university which does not offer any master's IT degree.

*Discussion:* Since the number of universities in Group-1 is much smaller than that in Group-2 at each level (e.g., bachelor's: Group 1 = 3; Group 2 = 35), it can be concluded that, in responding to the IT employment market trend, many universities have recognized the importance of offering MA-degree/SP-degrees to equip their IT graduates with knowledge that is "immediately" treasured by the tech sector.

### *RQ2: UoTs versus CUs*

Among the 39 offering universities, four are commonly known as UoTs: Queensland University of Technology, RMIT University, Swinburne University of Technology, and University of Technology Sydney. The remaining 35 universities are CUs.

We first counted and compared the number of MA-degrees and SP-degrees offered by UoTs and CUs. Note that the same MA-degree offered by a university may involve more than one major, e.g., Master of Computer Science (Software Engineering) and Master of Computer Science (Networking). To deal with this issue, when counting the number of MA-degrees, we took into account the number of different majors in the same MA-degree. Consider, for instance, an MA-degree with two majors: AI and cybersecurity. In this case, we counted it as two MA-degrees. This counting approach enables us to differentiate the "variety" of offering MA-degrees by a university in a more fine-grained manner.

Our counting and initial comparison results are shown in Table 4. At first glance, for both bachelor's and master's levels, the mean number of MA-degrees and the mean number of SP-degrees offered by UoTs are larger than the corresponding numbers offered by CUs. For example:

- Mean number of bachelor's MA-degrees: UoTs = 7.3; CUs = 4.4.
- Mean number of master's SP-degrees: UoTs = 3.0; CUs = 1.7.

Since our study covers the whole population (i.e., all IT degrees offered by all offering universities in Australia), the notion of significance or hypothesis testing (e.g., t-test) is irrelevant, because it only applies to a sample of data from a large population. As we are able to compute the true population parameters, hypothesis testing is not needed to infer the population data from a sample. Thus, we instead computed the relevant population effect size[5], showing the magnitude of a difference between two groups (e.g., UoTs and CUs). In essence, the effect size indicates the "practical" significance of a difference, i.e., whether an observed difference is "meaningful" in a real-world context.

Table 5 shows the result of effect size analysis. Here we used Cohen's d as the effect size measure.[6] This table shows that, when comparing with CUs, UoTs offer:

- more master's MA-degrees (Cohen's d = 0.51; medium difference); and
- significantly more bachelor's MA-degrees, bachelor's SP-degrees, and master's SP-degrees (Cohen's d $\geq$ 0.95; large difference).

*Discussion:* Being STEM-oriented, UoTs offer more career-focused education with an industry-aligned curriculum. Also, UoTs often build strong industry partnerships, offering degrees with built-in internships, real-world projects, and opportunities to work with emerging technologies. These arrangements increase graduate employability. Thus, it is understandable that UoTs offer more MA-degrees and SP-degrees with contemporary curricula bundled with advanced IT topics than CUs at both levels.

[4] Several universities have multiple campuses, even across different states of Australia. In our study, the geographic distribution of these universities is determined based on the locations of their head campuses. If a university does not have an explicit head campus, its first-established campus will be considered the "head" campus.

[5] Effect size analysis can be applied to the whole population or a sample of data.

[6] For Cohen's d, a value of about 0.2 is considered small, 0.5 is medium, and 0.8 or higher is large (Magnusson, 2025).

*RQ3: Go8 versus other universities*

There are eight Go8 universities and 31 other universities (i.e., non-Go8). Table 6 shows our counting and preliminary comparison. The mean numbers of bachelor's MA-degrees, master's MA-degrees, and master's SP-degrees offered by Go8 are higher than the corresponding numbers offered by other universities. For example:

- Mean number of bachelor's MA-degrees: Go8 = 6.0; other universities = 4.4.
- Mean number of master's SP-degrees: Go8 = 2.4; other universities = 1.7.

Table 7 shows the effect size analysis using Cohen's d. We observed two things from the table:

(a) At the bachelor's level, Go8 offer somewhat more MA-degrees (Cohen's d= 0.52) and marginally less (Cohen's d= −0.14) SP-degrees than other universities.
(b) At the master's level, Go8 offer more MA-degrees (Cohen's d=0.90) and somewhat more SP-degrees (Cohen's d=0.51) than other universities.

*Discussion:* At first sight, the above findings are puzzling. One may argue that Go8 are prestigious universities with more resources. Go8 attract significant funding through endowments, grants, and high tuition, thereby allowing them to offer more advanced and specialized programs. Following this argument, we would expect that Go8 will offer more MA-degrees and SP-degrees at both degree levels than other universities. However, although observation (b) above is in line with this expectation, observation (a) slightly falls short of the expectation.

A close examination of the counts taken from The University of Melbourne (UoM) may explain the unexpected observation (a). UoM, a member of Go8 and the top-ranked university in Australia (according to the 2025 Times Higher Education Ranking), introduced its new curriculum structure (called the "Melbourne Model") in 2008. This model replaced traditional bachelor's degrees with a two-tiered system, where students complete a broad bachelor's degree (e.g., arts or science) before moving on to a specialized (e.g., IT, engineering, and medicine) graduate-level study. Accordingly, UoM does not offer a specific bachelor's IT degree (hence, the number of GL-degrees, MA-degrees, and SP-degrees at the bachelor's levels are all zero). If students want to study IT at the bachelor's level, they have to enroll a Bachelor of Science (IT). UoM is the only university in Australia adopting this two-tiered system.

In view of the Melbourne Model, we consider UoM an outlier in our study. Since there are only eight Go8 universities, this outlier has a significant impact on the validity of our data analysis on Go8. To alleviate this problem, we repeated the effect size analysis for bachelor's IT degrees, but this time we ignored UoM, and only considered the remaining seven universities ("Go7") and all the other universities. Using this approach, when comparing Go7 with other universities, Cohen's d for bachelor's MA-degrees and SP-degrees are 0.84 (large difference) and −0.09 (very close to zero; effectively means no difference), respectively.

After replacing Go8 by Go7, we can now conclude that G07 offers at least the same mean number of (and often more) MA-degrees and SP-degrees as other universities at both degree levels.

*RQ4: Regional universities versus metropolitan universities*

A number of universities have multiple campuses in different geographical locations. Some of them have campuses in major cities and in regional areas. In this study, classifying a university as metropolitan or regional is made by reference to the location of their head campuses.[7] If their head campuses are located in a major (or the capital) city (see Table 1 for the capital/major city of each state/region) or a regional area, these universities are classified as metropolitan or regional, respectively. Based on Good Universities Guide (2025) and Regional Universities Network Australia (2025), we have compiled a list of 11 regional universities and another list of 28 metropolitan universities.

Refer to Table 8. The mean number of MA-degrees and SP-degrees at both levels offered by metropolitan universities are larger than those offered by regional universities (e.g., mean number of bachelor's MA-degrees: metropolitan universities=5.1; regional universities=3.6). The corresponding effect size analysis is shown in

[7] Some universities have multiple campuses but none of them is explicitly designated as the head campus. In this case, the original campus is considered the "head" campus.

Table 9, showing that across both levels, metropolitan universities generally offer slightly more (difference is only small to medium) MA-degrees and SP-degrees than regional universities.

*Discussion:* Initially, we speculated that regional universities would offer "obviously less" IT degrees than metropolitan universities. Our speculation was made based on the following reasons:

- Australia's regional universities generally have less student population than metropolitan universities, thereby restricting the formers' scale of operations and curriculum offerings.
- Most Australia's Indigenous people are living in rural/remote areas rather than in metropolitan cities (Australian Institute of Family Studies, 2025). Also, there exists a digital divide that hinders rural students' digital literacy and skills development (Turner, 2023). Consequently, Indigenous students are generally less interested in IT studies and careers (Australian Government Department of Education, 2025). Given this fact, regional universities would be less inclined to offer MA-degrees and SP-degrees.

Our effect size analysis in Table 9, however, shows a slightly different pattern. Our analysis shows that, although metropolitan universities offer more IT degrees than regional universities, the difference is only small to medium. Our close examination of the campus locations of regional universities seems to provide an explanation for this pattern. Among the 11 regional universities, 10 of them have campuses in at least one major city, despite their head campuses are located in regional areas. For example, the University of Newcastle (a regional university) has campuses in two major cities: Sydney and Melbourne. Intuitively, these regional universities with city campuses will have similar motivation to offer IT degrees as metropolitan universities due to the following:

- Regional universities with city campuses have a large number of metropolitan students in addition to rural students.
- Similar to students at metropolitan universities, metropolitan students at regional universities also have a great interest in studying IT, thus demanding their regional universities to offer IT training.

We conjecture that the above reasons will play a part in reducing the overall gap in the number of IT degrees offered between metropolitan and regional universities.

### *RQ5: Most popular SP-degrees*

We observed that some universities offer their SP-degrees which are essentially the same or very similar, but under slightly different names. An example is Bachelor of Data Science versus Bachelor of Data Analytics. To facilitate our analysis, we categorized similar specializations into one group, using common names such as data science and analytics.

Table 10 shows the total number of different SP-degrees at each level across all the 39 offering universities. If we consider the two levels together, the top four most popular specializations (in descending order of the total number of offerings) are: data science and analytics (=42), cybersecurity (=33), AI (=12), and software engineering (=8). These top four specializations together account for 78.3% (=(13+12)/46) and 85.9% (=(29+21)/71) of the total number of SP-degree offerings at the bachelor's and master's levels, respectively. Also, among these top four specializations, data science and analytics and cybersecurity are far more popular than the other two.

Among the 39 offering universities, 10 (=25.6%) of them offer at least three of the above top four specializations. These 10 universities include two Go8 universities: The Australian National University and Monash University.

*Discussion:* The above observation can be explained by a recent reporting by Monash University Malaysia (2025), which argues that "a diverse range of specializations exists within the expansive realm of computer science … which include cybersecurity, AI, software engineering, and data science, each possess a unique charm and offer many employment opportunities". We also observed that only one university offers a Master of Quantum Computing, which is one of the most significant recent technological advancements. We speculate that there are two reasons contributing to this limited offering:

- Quantum computing is new and evolving, suffering from various issues (e.g., scalability), Hence, quantum computing is yet to gain wide industry acceptance. In response to this, universities generally do not have a strong motivation to create a SP-degree in quantum computing.
- Creating and staffing a full SP-degree in quantum computing requires significant investment in expensive hardware equipment and recruitment of faculty staff knowledgeable in quantum physics, computer engineering, and software development. Not many universities can afford or are willing to make this investment.

### *RQ6: SP-degrees: Master's versus bachelor's*

Table 11 shows that the statistics of SP-degrees offered at both levels. Apparently, the total number of SP-degrees at the master's level (=71) is larger than that at the bachelor's level (=46). We then computed Cohen's d, which is equal to 0.09. Thus, Cohen's d indicates that the difference between the two means (bachelor's =1.2; master's =1.8) is in fact very minor.

*Discussion:* It is a common belief that bachelor's degrees are generally more broad and less specialized, whereas master's degrees are more specialized and provide in-depth training to students. Hence, one would expect that there will be more master's SP-degrees than bachelor's SP-degrees. However, this is not the case according to the computed Cohen's d. Nowadays, universities offer a comparable number of bachelor's SP-degrees as master's SP-degrees. This phenomenon may be attributed to the high industry demand for specialized IT knowledge and the growing complexity of tech sector.

### *RQ7: Most popular MA-degrees*

The issue of different names for similar majors also occurs across universities, as in the case of SP-degrees discussed above related to RQ5. To alleviate this issue, we applied the same treatment, i.e., using a common name for similar majors. All in all, we found 42 different majors of MA-degrees at each level.

Table 12 shows the number of offerings for each major at each level. If we consider both levels together, the top four most popular majors are: cybersecurity (=48), software engineering/development (=40), AI (=35), and data science and analytics (=28). Among the 39 offering universities, 15 (=38.5%) of them offer at least three of these four popular majors. Furthermore, among these 15 universities, four of them are Go8 universities: The Australian National University, The University of Sydney, The University of Queensland, and Monash University.

*Discussion:* There is a large array of different majors (=42) at each level offered across universities. This is beneficial to IT students because they have a large range of majors from which they can choose to fit into their own interests and career aspirations. We noted that the top four most popular majors are the same as the top four most popular specializations, although the relative popularity among these four majors/specializations is different. Also, similar to SP-degrees (see Table 10), some majors listed in Table 12 cover two different IT areas, such as computers and algorithms, and data science and AI. However, there are only one bachelor's and four master's SP-degrees covering two different IT areas (see the SP-degrees annotated with the dagger symbol "†" in Table 10). In contrast, the number of majors involving two different IT areas in MA-degrees as listed in Table 12 is much more: 11 at the bachelor's level and 13 at the master's level. One plausible reason is that, when students go to the extreme to study a SP-degree instead of an MA-degree, they really want to devote their concentration and effort to only one IT area. This strategy may increase the students' chance of being employed in their chosen IT area after graduation.

We also noted a similar phenomenon as in the case of SP-degrees — only two universities offer a major in quantum computing. This observation provides further support for our argument that Australian universities are generally not yet motivated to offer training in quantum computing as a major or a specialization.

### *RQ8: Degrees with a research component and/or an internship*

*Research component:* Here, we only counted those research components to be offered in one of the following ways: (a) a bachelor's honors year project[8], (b) a master's minor thesis[9], or (c) a "major" research project in a unit (compulsory or elective) either at the bachelor's or master's level.

Table 13 shows our counts at the university level. There are 84.6% of universities offering IT studies with a research component at the bachelor's and/or master's levels. Between the two levels, offering a research component is more popular at the master's level (74.4%) than at the bachelor's level (34.2%). Among the six (=39−33) universities without offering a research component in their IT studies, none of them is a Go8 university. This is in line with our expectation because all Go8 universities are research-intensive.

Next, we consider the counts at the degree level. Table 14 shows that each of the three types of master's IT degrees with a research component has a percentage at least double that of its counterpart at the bachelor's level (e.g., bachelor's MA-degrees with a research component: 25.5%; master's MA-degrees with a research component: 56.4%). Also, if we do not differentiate the different types, the overall percentages of all IT degrees with a research component at the bachelor's and master's levels are 23.9% (=(7+47+8)/(29+184+46)) and 57.3% (=(14+75+40)/(21+133+71)), respectively.

---

[8] In Australia, a bachelor's honors year project is an optional, one-year extension of an undergraduate degree where students pursue an independent research project under the guidance of an academic supervisor.

[9] In Australia, some master's studies by coursework involve a minor-thesis component.

*Discussion:* As stated above, 84.6% of universities offer IT studies with a research component. We see two possible reasons for this. First, the Australian Government has developed a national research workforce strategy (called Research Skills for an Innovative Future) to promote research training offered by universities (Department of Innovation Industry, Science and Research, 2011). Second, according to a survey by The Australian National University (2017), among the 29 SEEK's industry categories[10], the IT sector was ranked fourth in terms of the number of jobs demanding high levels of research skills. This industry demand in turn puts pressure on universities to incorporate more research elements into their IT curricula.

Based on Table 14, we computed that the overall percentage of all types of IT degrees with a research component at the master's level (57.3%) is more than doubled than that at the bachelor's level (23.9%). The rationale behind this finding is obvious. When compared with a bachelor's degree, a coursework master's degree is more likely to incorporate a research component (e.g., a minor thesis), because the latter aims to prepare graduates for leadership positions or further doctoral studies by acquiring specialized knowledge and advanced research skills.

*Internship:* Nowadays, different forms of work integrated learning (WIL) exist. Besides traditional industrial placements (or internships), Australian universities have also incorporated other forms of WIL (e.g., industry/community projects, simulations, and virtual placements) into their IT curricula. Our study focuses only on internships (related to RQ8 and RQ10), which offer hands-on experience to students within a host company. Often, an internship involves co-supervision by an academic and a workplace supervisor; the latter has a clearly defined role and responsibility to provide practical skills to the student undertaking work placement.

Table 15 shows the numbers and percentages of each type of IT degrees with an internship across the two levels. In this table, we counted the number of MA-degree by ignoring the number of different majors in the same MA-degree. This counting approach was based on our observation of the collected data that, if an MA-degree offers an internship option, this arrangement will apply to all the majors of this MA-degree. Thus, there was no need to differentiate the different majors of the same MA-degree.

When considering the three types of IT degrees together, internships are less common at the master's level (overall 20.5%) than at the bachelor's level (overall 44.3%). Now, we focus on the bachelor's level. Table 15 shows that the percentages of degrees with an internship across the three types are close, with a range between 42.6–45.5%. This is, however, not the case for the master's level. The percentage starts from 9.5% for GL-degrees, rises to 20.0% for MA-degrees, and finally reaches 23.9% for SP-degrees.

*Discussion:* Regarding the first observation that master's IT degrees have fewer internships, we conjecture two reasons. First, many master's students already have work experience, so the focus of their studies shift to offering more advanced academic knowledge instead of providing hands-on job experience. Second, some master's students have already gone through their internships in their past undergraduate studies. Hence, these students may prefer to enroll a master's degree whose curriculum is packed with more advanced on-campus training rather than enrolling one with an internship component. Regarding the second observation that the percentage of master's IT degrees with internships increases with the level of specialization (from GL-degrees through MA-degrees to SP-degrees), IMFS (2025) provides an explanation. Specialized master's IT programs are often designed to be more professionally focused, aiming at preparing work-ready graduates with industry-specific skills and experience. Hence, when compared with general master's IT programs, their specialized counterparts emphasize more on practical application, so that graduates can effectively perform their jobs in a narrow IT field with less on-the-job training.

### *RQ9: Degrees with a supporting infrastructure*

*Professorship:* There is an observed trend that universities prefer to recruit full professors for the long-term strategic growth of a discipline, instead of just filling a vacant position. This trend reflects the contribution of full professors towards shaping a school's academic profile by providing leadership in research, teaching, and recruitment of other academic staff (Levander et al., 2022). Inspired by this observation, we analyzed whether, for each IT major/specialization offered by a university, one or more full professors (thereafter simply referred to as "professors") with the relevant expertise are recruited for this major/specialization.

Our study only counted "regular" full professors. Visiting professors, emeritus professors, adjunct professors, honorary professors, and professorial fellows were excluded. Tables 10 and 12 show that some majors or specializations include two IT areas, e.g., a master's SP-degree in cybersecurity and AI. To facilitate our analysis here, for each "double-IT" major or specialization, we considered it two "separate" majors or specializations. For example, we considered the specialization "cybersecurity and AI" two "separate" specializations: one is cybersecurity and the other one is AI. Then, we checked whether a university has recruited professors whose research expertise covers cybersecurity or AI.[11]

[10] SEEK Limited is an Australian online employment marketplace for job listings.

For each university, we first computed the percentage of majors and specializations (considering the bachelor's and master's levels together) covered by its full professors' research expertise ("percentage of coverage"). Considering all the 39 offering universities together, the mean percentage of coverage is 44.6%, with a range of 0.0–100.0%, and a standard deviation of 31.6%. Next, we categorized all universities into two groups (Go8 and non-Go8) and recomputed the above summary statistics. The results are shown in Table 16.

*Discussion:* Considering all the 39 universities together, the mean percentage of coverage is 44.6%. Since this percentage is close to half, it provides further support for the above-mentioned trend that universities prefer to recruit professors whose research expertise matches to the disciplines offered. In Table 16, the statistics show that Go8 universities have a much higher mean percentage of coverage (72.0%) than non-Go8 universities (37.6%). Apparently, there are two possible facts leading to Go8's higher mean percentage of coverage: (Possible fact 1) Go8 universities do recruit more professors for the offered IT majors/specializations than non-Go8 universities. (Possible fact 2) The numbers of different IT majors/specializations offered by Go8 universities are much less than non-Go8 universities. Intuitively, a small number of majors/specializations offered by Go8 universities will easily give rise to their high percentage of coverage.[12] To examine this issue, we also computed the mean numbers of IT majors/specializations offered by Go8 and non-Go8 universities. The mean number of IT majors/specializations offered by Go8 is 9.4, which is larger than that by non-Go8 universities (whose mean number is 6.5). Thus, these mean numbers have confirmed that possible fact 2 is invalid. In other words, possible fact 1 applies.

There are two plausible reasons for fact 1. First, Go8 universities have more funding resources, thereby allowing them to recruit more professors to fit their teaching and research needs. Second, by virtue of their higher academic reputation, Go8 universities are more able to attract high-caliber academics to fill their professorship positions.

*Research institute or center:* Lindsay et al. (2002) reported that research-informed teaching improves students' learning experience and greater employability. Inspired by these studies, we analyzed the number of research institutes or centers[13] established in offering universities that are relevant to their MA-degrees or SP-degrees. In this analysis, we did not count informal and small research groups or clusters. Among all the 39 offering universities, the total number of research institutes established is 48. The mean number of research institutes per each university is 1.23 (=48/39), with a range [0, 6], and a standard deviation of 1.54.

Due to their prestigiousness, Go8 universities are able to attract more funding and, hence, are expected to have more financial resources to establish their research institutes. Under this rationale, we reperformed the above analysis separately for Go8 universities and non-Go8 universities. Table 17 shows the results. At first glance, the mean number of research institutes per each university of Go8 (=1.54) is larger than non-Go8 (=1.50). In effect size analysis, Cohen's d is 0.54, indicating only a medium difference.

*Discussion:* We computed the mean number of different IT areas associated with the MA-degrees and SP-degrees (at both bachelor's and master's levels) across all the 39 offering universities. The mean number is 7.03, which is much larger than the mean number of research institutes per each university (=1.23). This indicates that, overall, offering universities only spend their resources to establish research institutes for some highly selected (rather than all) IT areas to fit their own strategic research agendas.

Also, the Cohen's d value of 0.54 indicates that Go8 universities have established more research institutes than non-Go8 universities, but only with a medium difference. There are two plausible reasons contributing to this medium difference:

- Nowadays, although Go8 universities still attract a significant majority of research funding from the Australian Government (e.g., the Australian Research Council) and other non-profit sectors to establish/subsidize their research institutes (Parliament of Australia, 2022), non-Go8 universities have a more diversified research funding model by deriving a large percentage of their research incomes from industry and external partners (Innovative Research Universities, 2023). To some extent, this helps reduce the gap in research incomes between Go8 and non-Go8 universities.
- As discussed above, offering universities (both Go8 and non-Go8) only spend their resources to establish highly selected research institutes. By focusing on specific IT research areas, universities can achieve a critical mass of expertise and resources needed for creating higher research impacts and innovation. Thus, even Go8 universities have more resources, they may prefer to enlarge the scale of their existing research institutes and improve their supporting facilities and infrastructure, rather than establishing new research institutes.

---

[11] Consider, for example, a university offers neither an MA-degree majoring in AI nor a SP-degree in AI. Even if this university has recruited a professor in AI, this professor will not be counted for our analysis.

[12] Consider, for example, a hypothetical university which only offers one IT major or specialization. In this case, even if this university recruits only one professor with the relevant research expertise, its percentage of coverage will become 100%.

[13] To avoid verbosity, research institutes and research centers will be collectively referred to as "research institutes".

### *RQ10: Industrial involvement*

We focus on the following four types of industrial involvement for those universities offering MA-degrees and/or SP-degrees:

(a) Industry advisory board: It advises the university on external industry needs and connect the university with the IT industry sector.
(b) Curriculum co-design: This university-industry collaboration ensures that an IT curriculum will produce graduates who are ready for the IT workforce (Laundon et al., 2023).
(c) Co-teaching: This joint effort between academics and industry experts brings together two different teaching styles, backgrounds, and perspectives, which help students better relate to and engage with instructors (Busteed, 2025). To some extent, leveraging this collaboration also helps universities struggling to satisfy unfulfilled student demand for internship experience. In this study, we only considered those co-teaching arrangements that are "formally" incorporated in the curriculum. Ad hoc, irregular arrangements of industry seminars would be ignored.
(d) Internship: Here, we investigated the internship arrangement at the IT school (or university) level, i.e., the availability of internships in each type of IT degrees.

We noted that all the 39 universities offer at least one MA-degree or SP-degree. Table 18 shows the statistics related to industry advisory board, curriculum co-design, and co-teaching. Table 19 shows the numbers and percentages of universities offering at least one IT degree with an internship.

*Discussion:* Consider Table 18. Overall, the level of industry involvement related to types (a), (b), and (c) is fairly low. Relatively speaking, universities prefer to establish an industry advisory board (20.5%) than implementing curriculum co-design (7.7%) and co-teaching (0.0%). We speculate that it is because the effort associated with operating an industry advisory board is not high (from both the university's and the industry's perspectives), as it typically involves only a few meetings a year. On the other hand, the effort demanded by curriculum co-design and co-teaching is much higher, which hinders many universities from implementing them.

Next, we turn to Table 19. This table shows that, considering the bachelor's and master's levels together, more than half (59.0%) of universities offering internships in at least one of their IT degrees, indicating that universities generally consider that internships play a key role in training students and improving employability (Chillas et al., 2015). Considering Tables 18 and 19 together, it is clear that internships are far more popular than the other three types of industrial involvement.

## Further discussions and recommendations

Based on our observations and findings, we have the following three recommendations.

### *Technology-centered curriculum versus career-centered (or application-centered) curriculum*

Technology is not an isolated island. Rather, it is ultimately integrated into and shapes the various kinds of sociotechnical systems in organizations. Despite this, we see from Tables 10 and 12 that the design of most SP-degrees and MA-degree majors primarily centers around the technology rather than its related careers or applications. Our observation is supported by Herbert et al. (2013), who argued that, although universities often use career outcomes to market their IT degrees to potential students, there is little evidence that career-driven outcomes have in fact been embedded into IT curricula.

Consider, for example, the financial technology (FinTech) career. Since the last decade, FinTech has been experiencing tremendous growth due to significant investment and its disruptive impact on traditional finance through a wide range of supporting technologies (Kou & Lu, 2025). Examples of these technologies are AI, ML, virtual reality, data science and analytics, cybersecurity, and quantum computing.

Despite the prosperous employment aspect of FinTech, Tables 10 and 12 show that: (a) none of the offering universities offers a SP-degree of FinTech or an MA-degree majoring in FinTech, and (b) many existing SP-degrees and MA-degree majors focus on only one of the FinTech supporting technologies, such as Bachelor of AI, Master of ML, Bachelor of Data Science and Analytics, and Master of IT (Cybersecurity). Should graduates of the above degrees decide to enter the FinTech workspace upon completing their studies, they will find that their degrees do not cover all the technical knowledge needed by them to successfully start their FinTech career.

One may argue that some Australian universities offer their specialized FinTech degrees (e.g., the University of New South Wales offers an online Master of Financial Technology). However, these degrees are offered by Business/Finance schools rather than by IT schools. As such, these "business-oriented" FinTech degrees are designed to focus more on the user perspective of FinTech, rather than from the developer perspective. Thus, these FinTech degrees are not an option for those prospective students looking for an IT degree which prepares them to work in the FinTech space as a FinTech application developer (rather than a user). In view of the above issue, we recommend IT schools should consider offering their SP-degrees and MA-degree majors (e.g., Bachelor/Master of IT (Financial Technology)) which center around an IT career (e.g., FinTech).

*Co-teaching involving two universities*

During our data collection, we were not aware of any "formal" arrangement for co-teaching IT units involving instructors from two different universities — a partnership where both universities' faculty members co-instruct a unit to create a shared, equitable learning environment for students. There are two approaches of this co-teaching model, where the two instructors from different universities: (a) share the teaching load, with each of them responsible for teaching a class separately, or (b) co-teach some or all classes together. This co-teaching model may also involve the two instructors to co-design the syllabus, assessments, and grading rubrics.

Haag et al. (2023) observed that, from the students' perspective, this co-teaching model has the following merits:

- A higher level of class engagement: Students are more actively engaged, motivated, and invested in their learning.
- Increased learning outcomes: This is achieved by exposing students to diverse perspectives and teaching styles.
- Acceptance of knowledge limit: For approach (b) above, students will pay close attention to the interaction between the two instructors. On some occasions, students may even observe that one instructor (or scholar) questions or corrects the other. Students will then learn from this observation that there are limits to any scholar's knowledge. This will in turn become a model of the students' own role and behavior, where they take an acceptance of limits to knowledge with them beyond the co-taught unit.

In view of the above merits, we therefore recommend IT schools of different universities considering the feasibility of implementing the co-teaching model for the benefits of their students.

*Co-creating syllabi with students*

Similar to co-teaching across universities as discussed above, we were not aware of any offering universities that involve students in co-designing an IT syllabus (or even the entire curriculum). In this model, we view students as our "partners". More specifically, we engage with students before, during, or after a unit to redesign its syllabus or co-design a new unit (Hubbard et al., 2017). The University of Queensland (n.d.) argues that this model offers the following benefits:

- Skill development: Co-designing syllabi with students allow them to develop critical thinking, problem solving, and metacognitive skills.
- Deeper understanding: Students form a stronger connection to the teaching materials, leading to better comprehension and retention.
- Improved learning outcomes: A co-designed syllabus is often more aligned with real-world needs and potentially leads to better academic progress.
- Personalized learning journey: Through contributing their unique perspectives, students are able to shape the learning contents to fit into their own interests and goals.

Although co-designing syllabi with students requires extra effort for collaboration and a different mindset compared with traditional methods, it is worth spending the extra effort in view of the above merits .

## Study limitations

Ideally, all the data should be collected within a very short time period for more accurate comparison and analysis. However, this was not feasible due to the large number of participant universities and a substantial number of online web pages from which data were collected. In this study, data collection spanned about two months (October–November 2025) to complete. In principle, though unlikely, some changes could have happened in the online web pages amidst our data collection work. However, subject to our available resources, we have already made our best effort to shorten the data collection period, with a view to minimizing any effect that may invalidate the results of our study.

Also, our study was conducted based on the online data collected from the participant university's websites. It was possible that the contents of some of these websites were not up to date. Nevertheless, even if there were some web pages which were not updated, our results still paint an overall picture of the status quo of the Australian universities in offering GL-degrees, MA-degrees, and SP-degrees to their students.

## Conclusion and future work

In this paper, we have discussed our recent census study to investigate the status quo of offering the three different types of IT degrees (GL-degree, MA-degree, and SP-degree) at both bachelor's and master's levels. Our census study covered all Australian universities, except three that do not offer any IT study.

Some major findings of our study are: (1) many universities now offer MA-degrees and SP-degrees; (2) UoTs offer more MA-degrees and SP-degrees than CUs; (3) Go8 universities (except the University of Melbourne) offer at least the same mean number of (and often more) MA-degrees and SP-degrees as non-Go8 universities; (4) regional universities with city campuses offer more IT degrees than regional universities without city campuses; (5) the top four most popular MA-degrees and SP-degrees are related to data science and analytics, AI, cybersecurity, and software engineering; (6) universities offer a comparable number of bachelor's SP-degrees as master's SP-degrees; (7) the majority of universities offer IT studies with a research component, and a research component is more common at the master's level than the bachelor's level; (8) master's IT degrees have fewer internship than their bachelor's counterparts, and the percentage of master's IT degrees with internships increases with the level of specialization (from GL-degrees through MA-degrees to SP-degrees); (9) universities prefer to recruit "full" professors whose research expertise matches to the IT disciplines offered, and Go8 universities recruit more "full" professors for the offered IT majors/specializations than non-Go8 universities; (10) universities only spend their resources to establish research institutes for some highly selected IT areas to fit their own strategic research agendas; and (11) among the various types of industrial involvement, the establishment of an industry advisory board in universities is the most popular option.

Based on the above findings, we have made three recommendations for the benefits of the students (in terms of learning outcomes and graduate employability): creating career-centered IT curricula; co-teaching involving different universities; and co-creating syllabi with students.

## Declaration of conflicting interests

The author(s) declared no potential conflicts of interest with respect to the research, authorship, and/or publication of this article.

## Funding

The author(s) received no financial support for the research, authorship, and/or publication of this article.

## Data availability statement

The dataset for this study is available upon reasonable request.

**Table 1.** Geographical Locations of Offering Universities.

| Region | Capital/major city | No. (%) of offering universities in each region |
|---|---|---|
| Australian Capital Territory (ACT) | Canberra | 2 (5.1%) |
| New South Wales (NSW) | Sydney | 11 (28.2%) |
| Northern Territory (NT) | Darwin | 1 (2.7%) |
| Queensland (QLD) | Brisbane | 8 (20.5%) |
| South Australia (SA) | Adelaide | 2 (5.1%) |
| Tasmania (TAS) | Hobart | 1 (2.7%) |
| Victoria (VIC) | Melbourne | 9 (23.1%) |
| Western Australia (WA) | Perth | 5 (12.8%) |

**Table 2.** Numbers and Percentages of Offering Universities with respect to Each Type of IT Degrees.

| Degree level | Number (%) of universities offering | | |
|---|---|---|---|
| | GL-degree | MA-degree | SP-degree |
| Bachelor's | 19 (48.7%) | 31 (79.5%) | 22 (56.4%) |
| Master's | 17 (43.6%) | 29 (74.4%) | 31 (79.5%) |

**Table 3.** Numbers and Percentages of Universities Offering Only GL-Degree or Otherwise.

| Degree level | Number (%) of universities offering | | Total |
|---|---|---|---|
| | Group 1: only GL-degree | Group 2: MA-degree and/or SP-degree | |
| Bachelor's | 3 (7.9%) | 35 (92.1%) | 38 (100%) |
| Master's | 2 (5.3%) | 36 (94.7%) | 38 (100%) |

**Table 5.** Effect Sizes: UoTs versus CUs.

| Degree level | Cohen's *d* (UoTs versus CUs) | |
|---|---|---|
| | MA-degrees† | SP-degrees |
| Bachelor's | 0.95 (large difference) | 1.34 (large difference) |
| Master's | 0.51 (medium difference) | 0.95 (large difference) |

(†) Same counting approach as used in RQ2.
Note: A positive Gohen's *d* means that UoTs offer more degrees of the specific type than CUs.

**Table 7.** Effect Sizes: Go8 versus Other Universities.

| Degree level | Cohen's *d* (Go8 versus others) | |
|---|---|---|
| | MA-degrees† | SP-degrees |
| Bachelor's | 0.52 (medium difference) | – 0.14 (small difference) |
| Master's | 0.90 (large difference) | 0.51 (medium difference) |

(†) Same counting approach as used in RQ2.
Note: A positive (or negative) Gohen's *d* means that Go8 offer more (or less) degrees of the specific type than other universities.

**Table 9.** Effect Sizes: Metropolitan versus Regional Universities.

| Degree level | Cohen's *d* (metropolitan versus regional universities) | |
|---|---|---|
| | MA-degrees [†] | SP-degrees |
| Bachelor's | 0.48 (medium difference) | 0.64 (medium difference) |
| Master's | 0.45 (small to medium difference) | 0.36 (small to medium difference) |

(†) Same counting approach as used in RQ2.
Note: A positive Gohen's *d* means that metropolitan universities offer more degrees of the specific type than regional universities.

**Table 10.** Number of Different SP-Degrees.

| Bachelor's | | Master's | |
|---|---|---|---|
| SP-degrees | No. of offerings | SP-degrees | No. of offerings |
| Data Science & Analytics | 13 | Data Science & Analytics | 29 |
| Cybersecurity | 12 | Cybersecurity | 21 |
| Software Engineering | 6 | AI | 9 |
| Computer Games | 5 | Software Engineering | 2 |
| Computer Animation | 3 | Health Bioinformatics & Analysis | 2 |
| AI | 3 | Cybersecurity & AI [†] | 1 |
| Computers & Algorithms [†] | 1 | AI & ML [†] | 1 |
| User Experience (UX) Design | 1 | Networking | 1 |
| Digital Design | 1 | Cloud Computing | 1 |
| Digital Transformation | 1 | Data Engineering | 1 |
| Total | 46 | ML & Computer Vision [†] | 1 |
| | | Data Science & AI [†] | 1 |
| | | Quantum Computing | 1 |
| | | Total | 71 |

(†) SP-degrees involving two specialized IT areas.

**Table 11.** SP-Degrees: Bachelor's versus Master's.

| Degree level | SP-degrees | | |
|---|---|---|---|
| | Total number | Mean number | Standard deviation |
| Bachelor's | 46 | 1.2 | 4.3 |
| Master's | 71 | 1.8 | 8.7 |

**Table 12.** Number of Offerings of Each Major in MA-Degrees.

| Bachelor's | | Master's | |
|---|---|---|---|
| No. of offerings | Majors | No. of offerings | Majors |
| 28 | Cybersecurity | 20 | Cybersecurity |
| 22 | Software Engineering/Development | 18 | Software Engineering/Development |
| 18 | AI | 17 | AI |
| 14 | Data Science & Analytics | 12 | Data Science & Analytics |
| 8 | Computer Games, Networking, Networking & Cybersecurity† | 9 | Networking |
| 7 | Computer Science | 6 | Human-Centered Computing |
| 6 | Data Science & AI, Web/Mobile App Development† | 3 | Cloud Computing, Computer Science, Data Science & AI†, Networking & Cybersecurity†, Systems Analysis |
| 5 | Enterprise Computing | 2 | Applied Computing, Enterprise Computing, IoT, ML, Multimedia Computing |
| 4 | Programming, Systems Analysis, Systems Security | 1 | AI & ML†, Algorithms, Blockchain, Cloud & Distributed Computing†, Cloud Computing & IoT†, Computer Systems, Computers & Algorithms†, Computing Technology, Data Science & ML†, Database Systems, Distributed Computing, Emerging Technologies, Health Informatics, Information Technology, Mobile & Cloud Computing†, Mobile App Development, Mobile Computing & IoT†, Network & Cloud Computing†, Networking & Distributed Systems†, Programming & Distributed Computing†, Quantum Computing, Space Science, Spatial Information, Virtual Reality, Web Systems, Web/Mobile App Development† |
| 3 | Human-Centered Computing | | |
| 2 | Cloud Computing, Computers & Algorithms†, Data Science & IoT†, Database Systems, Interactive Design, IoT, ML, Network & Cloud Computing†, Quantum Computing, Software Technology | | |
| 1 | AI & Autonomous Systems†, AI & ML†, Computational Mathematics, Computer Animation, Computer Systems & Robotics†, Digital Forensics, Digital Innovation, Embedded Systems, Game/Mobile App Development†, Health Informatics, Intelligent Systems, Information Technology, Mobile & Cloud Computing†, Multimedia Computing, Robotics, Systems & Architecture†, UX Design | | |

(†) Majors involving two different IT areas.

**Table 13.** Numbers and Percentages of Universities Offering IT Studies with a Research Component.

| | Bachelor's or master's | Bachelor's | Master's |
|---|---|---|---|
| No. (percentage) of universities offering IT studies with a research component | 33 (84.6%) | 13 (34.2%)[†] | 29 (74.4%) |

(†) Counting at the bachelor's level did not involve The University of Melbourne, because it does not offer any IT study at the bachelor's level.

**Table 14.** Number and Percentages of Each Type of IT Degrees with a Research Component.

| | Bachelor's | | | Master's | | |
|---|---|---|---|---|---|---|
| | GL-Degrees | MA-degrees | SP-degrees | GL-Degrees | MA-degrees | SP-degrees |
| No. of each type of degree offered | 29 | 184 | 46 | 21 | 133 | 71 |
| No. of degrees with a research component | 7 | 47 | 8 | 14 | 75 | 40 |
| Percentage of degrees with a research component | 24.1% | 25.5% | 17.4% | 66.7% | 56.4% | 56.3% |

**Table 15.** Numbers and Percentages of Each Type of IT Degrees with and without an Internship.

| | Bachelor's | | | Master's | | |
|---|---|---|---|---|---|---|
| | No. (%) of degrees with internship | No. (%) of degrees without internship | Total no. (%) of degrees | No. (%) of degrees with internship | No. (%) of degrees without internship | Total no. (%) of degrees |
| GL-degree | 13 (44.8%) | 16 (55.2%) | 29 (100%) | 2 (9.5%) | 19 (90.5%) | 21 (100%) |
| MA-degree | 21 (42.6%) | 28 (59.2%) | 49 (100%) | 7 (20.0%) | 28 (80.0%) | 35 (100%) |
| SP-degree | 20 (45.5%) | 24 (54.5%) | 44 (100%) | 17 (23.9%) | 54 (76.1%) | 71 (100%) |
| Total (%) | 54 (44.3%) | 68 (55.7%) | 122 (100%) | 26 (20.5%) | 101 (79.5%) | 127 (100%) |

**Table 16.** Summary Statistics on the Percentages of Coverage between Go8 and Non-G08 Universities.

| Group of universities | Percentage of coverage | | |
|---|---|---|---|
| | Mean | Range | Standard deviation |
| Go8 | 72.0% | 16.7–100.0% | 25.8% |
| Non-Go8 | 37.6% | 0.0–100.0% | 29.0% |

**Table 17.** Summary Statistics on the Number of Research Institutes in Go8 and Non-G08 universities.

| Group of universities | Number of research institutes | | |
|---|---|---|---|
| | Mean | Range | Standard deviation |
| Go8 | 1.88 | 0–4 | 1.54 |
| Non-Go8 | 1.06 | 0–6 | 1.50 |

**Table 18**. Numbers and Percentages of Universities Involving an IT Industry Advisory Board, Curriculum Co-Design, and Co-Teaching.

| | Industry advisory board | Curriculum co-design | Co-teaching |
|---|---|---|---|
| No. (percentage) of offering universities | 8 (20.5%) | 3 (7.7%) | 0 (0.0%) |

**Table 19.** Numbers and Percentages of Universities Offering IT Studies with an Internship

| | Bachelor's or master's | Bachelor's | Master's |
|---|---|---|---|
| No. (percentage) of universities offering IT studies with an internship | 23 (59.0%) | 19 (48.7%)[†] | 12 (30.8%) |

(†) Counting at the bachelor's level did not involve The University of Melbourne, because it does not offer any IT study at the bachelor's level.

**Table 4.** Descriptive Statistics: UoTs versus CUs.

| | Universities of technology (UoTs) | | | | | | Comprehensive universities (CUs) | | | | | |
|---|---|---|---|---|---|---|---|---|---|---|---|---|
| | Bachelor's | | | Master's | | | Bachelor's | | | Master's | | |
| | No. of MA-degrees* | No. of MA-degrees† | No. of SP-degrees | No. of MA-degrees* | No. of MA-degrees† | No. of SP-degrees | No. of MA-degrees* | No. of MA-degrees† | No. of SP-degrees | No. of MA-degrees* | No. of MA-degrees† | No. of SP-degrees |
| Range | [1, 2] | [5, 10] | [1, 5] | [1, 1] | [3, 6] | [2, 4] | [0, 3] | [0, 10] | [0, 5] | [0, 3] | [0, 11] | [0, 6] |
| Mean | 1.8 | 7.3 | 2.8 | 1.0 | 4.8 | 3.0 | 1.2 | 4.4 | 1.0 | 0.9 | 3.3 | 1.7 |
| Standard deviation | 0.43 | 2.28 | 1.48 | 0.00 | 1.09 | 0.71 | 0.82 | 3.12 | 1.33 | 0.71 | 3.06 | 1.41 |

(*) Count an MA-degree with different majors as "one single" MA-degree.
(†) Count an MA-degree with different majors as "different" MA-degrees.

**Table 6.** Descriptive Statistic: Go8 versus Other Universities.

| | Go8 | | | | | | Other universities | | | | | |
|---|---|---|---|---|---|---|---|---|---|---|---|---|
| | Bachelor's | | | Master's | | | Bachelor's | | | Master's | | |
| | No. of MA-degrees* | No. of MA-degrees† | No. of SP-degrees | No. of MA-degrees* | No. of MA-degrees† | No. of SP-degrees | No. of MA-degrees* | No. of MA-degrees† | No. of SP-degrees | No. of MA-degrees* | No. of MA-degrees† | No. of SP-degrees |
| Range | [0, 2] | [0, 10] | [0, 4] | [0, 2] | [0, 11] | [1, 6] | [0, 3] | [0, 10] | [0, 5] | [0, 3] | [0, 8] | [0, 4] |
| Mean | 1.5 | 6.0 | 1.0 | 1.1 | 5.4 | 2.4 | 1.2 | 4.4 | 1.2 | 0.8 | 2.9 | 1.7 |
| Standard deviation | 0.71 | 2.96 | 1.32 | 0.60 | 3.74 | 1.58 | 0.82 | 3.13 | 1.47 | 0.68 | 2.48 | 1.33 |

(*) Count an MA-degree with different majors as "one single" MA-degree.
(†) Count an MA-degree with different majors as "different" MA-degrees.

**Table 8.** Descriptive Statistics: Metropolitan versus Regional Universities.

| | Metropolitan universities | | | | | | Regional universities | | | | | |
|---|---|---|---|---|---|---|---|---|---|---|---|---|
| | Bachelor's | | | Master's | | | Bachelor's | | | Master's | | |
| | No. of MA-degrees* | No. of MA-degrees† | No. of SP-degrees | No. of MA-degrees* | No. of MA-degrees† | No. of SP-degrees | No. of MA-degrees* | No. of MA-degrees† | No. of SP-degrees | No. of MA-degrees* | No. of MA-degrees† | No. of SP-degrees |
| Range | [0, 2] | [0, 10] | [0, 5] | [0, 2] | [0, 11] | [0, 6] | [0, 3] | [0, 7] | [0, 2] | [0, 3] | [0, 7] | [0, 4] |
| Mean | 1.3 | 5.1 | 1.4 | 0.9 | 3.8 | 2.0 | 1.2 | 3.6 | 0.5 | 0.9 | 2.5 | 1.5 |
| Standard deviation | 0.75 | 3.22 | 1.59 | 0.56 | 3.03 | 1.40 | 0.94 | 2.74 | 0.66 | 0.90 | 2.54 | 1.37 |

(*) Count an MA-degree with different majors as "one single" MA-degree.
(†) Count an MA-degree with different majors as "different" MA-degrees.